\documentclass[prd,amsmath,amssymb,superscriptaddress,floatfix,nofootinbib,10pt]{revtex4}
\usepackage{times}
\usepackage{amssymb,amsbsy,amsmath,amsfonts}
\usepackage{graphicx}
\usepackage{float}
\usepackage{color}
\usepackage{morefloats}
\usepackage{rotating}
\usepackage{srcltx}
\usepackage{slashed}
\usepackage{multirow}
\usepackage{verbatim}
\usepackage{hyperref}
\usepackage{tabularx}
\usepackage{bm}
\allowdisplaybreaks[4]

\usepackage{bbding}
\usepackage{threeparttable}

\DeclareUnicodeCharacter{2212}{\ensuremath{-}}

\newcommand{\PreserveBackslash}[1]{\let\temp=\\#1\let\\=\temp}
\newcolumntype{C}[1]{>{\PreserveBackslash\centering}p{#1}}
\newcolumntype{R}[1]{>{\PreserveBackslash\raggedleft}p{#1}}
\newcolumntype{L}[1]{>{\PreserveBackslash\raggedright}p{#1}}

\begin{document}

\title{Searching for signals  of an  exotic $I=1,J^P=2^+$ state of $D^*K^*$  nature and the structure of the $P_c(4312)$ in the $\Lambda_b\to \Sigma_c^{++} D^- K^-$ reaction}

\author{Jing Song}
\affiliation{School of Physics, Beihang University, Beijing, 102206, China}
\affiliation{Departamento de Física Teórica and IFIC, Centro Mixto Universidad de Valencia-CSIC Institutos de Investigación de Paterna, 46071 Valencia, Spain}

\author{Zi-Ying Yang}
\affiliation{School of Physics, Beihang University, Beijing, 102206, China}
\affiliation{Departamento de Física Teórica and IFIC, Centro Mixto Universidad de Valencia-CSIC Institutos de Investigación de Paterna, 46071 Valencia, Spain}

\author{ Eulogio Oset}
\email[]{oset@ific.uv.es}
\affiliation{Departamento de Física Teórica and IFIC, Centro Mixto Universidad de Valencia-CSIC Institutos de Investigación de Paterna, 46071 Valencia, Spain}
\affiliation{Department of Physics, Guangxi Normal University, Guilin 541004, China}

\begin{abstract}
This paper investigates the decay process \(\Lambda_b \to \Sigma_c^{++} D^- K^-\) with the objective of finding a predicted molecular state with isospin \(I=1\),   \(J^P=2^+\) of $D^* K^*$ nature, plus finding support for the $P_c(4312)$ state as made out of $\Sigma_c \bar{D}$.  The mass distribution of the \(D^- K^-\) system shows distinct features as a consequence of the existing of this $2^+$ state, while the \(\Sigma_c \bar{D}\) distribution exhibits a significant peak near the threshold, much bigger than phase space expectations, which is linked to our assumed $\Sigma_c \bar{D}$ nature of the $P_c(4312)$ state below the $\Sigma_c \bar{D}$ threshold. The reaction has been measured at LHCb Collaboration, but only the branching ratio is measured. The present study shows that much valuable information can be obtained about the predicted $2^+$  $T_{c\bar{s}}(2834)$ of $D^* K^*$ nature and the $P_c(4312)$ states from the measurements of the mass distributions in this reaction, which will be accessible in the planned updates of LHCb. 
\end{abstract}


\maketitle

\section{Introduction}
The discovery of exotic meson and baryon states, challenging the wisdom of mesons as $q \bar{q}$ states and baryon as made from three quarks, has triggered a great boost to hadron physics, and in particular to hadron spectroscopy, and many review papers have dealt with this issue 
~\cite{Chen:2016qju,Lebed:2016hpi,Esposito:2016noz,Guo:2017jvc,Ali:2017jda,Klempt:2007cp,Klempt:2009pi,Brambilla:2010cs,Olsen:2014qna,Oset:2016lyh,
Chen:2016spr,Hosaka:2016pey,Dong:2017gaw,Olsen:2017bmm}.
In the present work we shall look into two such states: one of them is a state predicted theoretically from the $D^* K^*$ and $D_s^* \rho$ interaction with isospin $I=1$ and $J^P=2^+$~\cite{Molina:2010tx}, and the other one is the well known $P_c(4312)$ state, discovered by the LHCb collaboration in Ref.~\cite{LHCb:2019kea} after missing it in the first run where other pentaquark states were reported~\cite{LHCb:2015yax}.

    In Ref.~\cite{Molina:2010tx}, among other states, described or predicted, three states were predicted with $I=1$ and $J^P=0^+,~1^+,~2^+$, coming from the $D^* K^*$ interaction in $S-$wave. A candidate for the first state with $0^+$, now called $T_{c\bar{s}}(2900)$, was observed by the LHCb collaboration~\cite{LHCb:2020bls,LHCb:2020pxc}. An update of the work of ~\cite{Molina:2010tx} was done in Ref.~\cite{Molina:2022jcd} to the light of the LHCb observation and more accurate predictions were done for the states of $0^+,~1^+,~2^+$. The first two states were very close to threshold, but the The first two states were very close to threshold, but the $2^+$ 
    state, at 2834 MeV  was very bound, by about 86 MeV. This is not surprising, since in all studies of the vector meson-vector meson interaction leading to bound states, initiated in Refs.~\cite{Molina:2008jw,Geng:2008gx}, the $2^+$ state is always more bound. This large binding of the $2^+$ state is corroborated by other independent studies~\cite{Duan:2023lcj}, were also a large binding of 114-140 MeV is found. More works in this direction can be seen in Refs.~\cite{Lyu:2024wxa,Lyu:2023aqn,Lyu:2023ppb,Duan:2023qsg,Yue:2022mnf}.  Work on these systems using molecular components with the interaction at the quark level are also done in Ref.~\cite{Ortega:2023azl}, where the
    $2^+$  
     state also appears and more bound than the $0^+$. The molecular picture for the $T_{c\bar{s}}(2900)$ is also supported in sum rule calculations~\cite{Agaev:2022eyk}.  There is hence a broad support for the existence of the $I=1, ~J^P=2^+$ state of  $D^* K^*$ nature. The state will naturally decay into $DK$, but in angular momentum  $L=2$ ~\cite{Molina:2022jcd}. This is the basic ingredient that we shall take into consideration for the analysis here. 
    
       Concerning the pentaquark states~\cite{LHCb:2020bls,LHCb:2020pxc}, it is appropriate to recall that, using similar dynamics as in ~\cite{Molina:2010tx}, realistic predictions of pentaquark states as molecular states were done in Refs.~\cite{Wu:2010jy,Wu:2010vk}  prior to their discovery. Other works followed in Refs.~\cite{Wu:2010jy,Wu:2010vk,Wang:2011rga,Sun:2012zzd,Wu:2012md,Xiao:2013yca,Li:2014gra,Chen:2015loa,Karliner:2015ina}. A common 
picture for the $P_c(4312)$ state that we address here is that it corresponds to a molecular state of $\Sigma_c \bar{D}$~\cite{Xiao:2013yca,Du:2020bqj,Du:2019pij,Uchino:2015uha,Chen:2019bip,Chen:2019asm,He:2019ify,Guo:2019kdc,Xiao:2019mvs,Zhang:2019xtu,Sakai:2019qph,Wang:2019hyc,Yamaguchi:2019seo,Xu:2019zme,PavonValderrama:2019nbk,Peng:2019wys,Lin:2019qiv,He:2019rva,Wang:2019ato,Wang:2019spc,Xu:2020gjl,Chen:2020pac,Dong:2020nwk,Kuang:2020bnk,Xiao:2020frg,Xu:2020flp,Azizi:2020ogm,Liu:2020hcv,Dong:2021juy,Peng:2021hkr,Du:2021fmf,Zou:2021sha,Phumphan:2021tta,Chen:2021xlu,Shi:2021wyt,Dong:2021rpi,Yalikun:2021bfm,Chen:2021cfl,Xing:2022ijm,Li:2023zag,Ozdem:2024jty}. This is not the only view on its nature since other works have provided a picture based on hadrocharmonium dynamics~\cite{Eides:2019tgv}, a double triangle cusp effect~\cite{Nakamura:2021qvy}, or a compact pentaquark structure~\cite{Wang:2019got,Cheng:2019obk,Pimikov:2019dyr,Xu:2019fjt,Chen:2019thk,Nakamura:2021qvy,Ruangyoo:2021aoi,Nakamura:2021dix,Garcilazo:2022kra,Garcilazo:2022edi,Wang:2024unj}. As we can see, although the majority of the works support the molecular picture, there is no consensus regarding the nature of the 
$P_c(4312)$ state. This is why an experiment linking this state to $\Sigma_c \bar{D}$ would shed valuable light concerning this issue.  Note that, so far, the $P_c(4312)$ state is observed in the $J/\psi p$ invariant mass distribution in the $\Lambda_b^0\to J/\psi pK^−$ reaction, and  nothing can make us think from this source that the state is linked to $\Sigma_c \bar{D}$. 

   One opportunity to make this test is offered by the recent LHCb reaction $\Lambda_b\to \Sigma_c^{++} D^- K^-$~\cite{LHCb:2024fel}, by looking at the $\Sigma_c^{++} D^- $ invariant mass close to threshold.  The $P_c(4312)$ state lies only 9 MeV below the $\Sigma_c^{++} D^- $ threshold and the $\Lambda_b\to \Sigma_c^{++} D^- K^-$ can proceed in $S-$wave. Thus, we expect a large enhancement of the $\Sigma_c^{++} D^- $ mass distribution around threshold due to the presence of the resonance just below. Of course, this is only in the case that the $P_c(4312)$ resonance largely couples to $\Sigma_c \bar{D}$. Should the $P_c(4312)$ be a genuine state with no links to  $\Sigma_c \bar{D}$, then we would not observe any enhancement. The amount of he enhancement observed with respect to phase space should be a measure of the coupling of the $P_c(4312)$ to the $\Sigma_c \bar{D}$ component. We will make predictions for this invariant mass distribution based on the molecular picture, which can be tested with experiment when it is implemented in the future. So far, only the branching fraction for this reaction relative to $\Lambda_b^0\to\Lambda_c^+D^0K^-$ has been measured~\cite{LHCb:2024fel}, but future runs in planned updates of LHCb should make this attainable. 
   
   It is interesting to note that the same reaction,  $\Sigma_c^{++} D^- K^-$, offers also the possibility to look at the $D^- K^-$ mass distribution, where one could observe the $I=1,~ J^P=2^+$ state of Ref. ~\cite{Molina:2010tx}. We anticipate clear signals based on the analogous reaction $B^+\to D^{*+} D^{+} K^+$, also measured at LHCb~\cite{LHCb:2024vfz}, where one signal appeared at 2834 MeV, precisely the mass where the $2^+$ state was predicted in the updated work of Ref.~\cite{Molina:2022jcd}. In Ref.~\cite{Lyu:2024zdo}  it was shown that the signal was compatible with the existence
of this $2^+$ state based on the $D^+ K^+$ mass distribution, and calculations of moments of mass-angle distributions were done, which magnify the signal and give information about the spin of the state. We shall do the same here. We will assume that the strength of $\Lambda_b\to \Sigma_c^{++} D^- K^-$ the decay is similar to that of $B^+\to D^{*+} D^{+} K^+$, since the weak decay follows the same pattern, and then we will make predictions for both the mass distributions and the moments, which should allow us determine the spin of the system.   These two aspects of the reaction, giving information on two very different states should serve as a motivation to look at these distributions whenever higher statistic samples become available.

\section{Formalism}

\begin{figure}[H]
  \centering
   \includegraphics[width=0.4\textwidth]{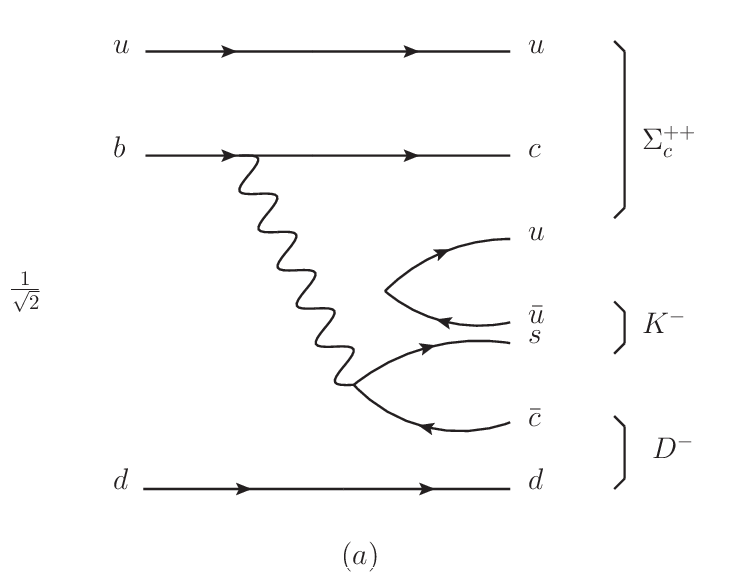}~~
    \includegraphics[width=0.45\textwidth]{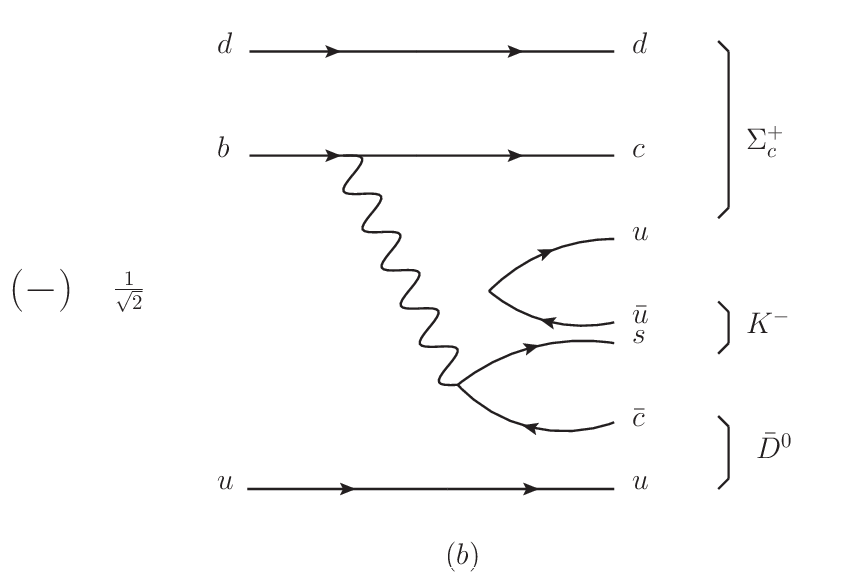}
  \caption{Feynman diagrams for the reaction \( \Lambda_b \to \Sigma_c^{++} D^- K^- \).}
   \label{feynDiag_Wission}
\end{figure}

To describe the mass distribution of the \( D^- K^- \) and \( \Sigma_c \bar{D} \) systems in the decay \( \Lambda_b \to \Sigma_c^{++} D^- K^- \), we will employ two key mechanisms.
Initially, we focus on the Feynman diagrams corresponding to the decay of the \( \Lambda_b \) baryon into \( \Sigma_c^{++} D^- K^- \).
Using the quark model decompositions~\cite{Capstick:1986ter,Roberts:2007ni} \( \Lambda_b = \frac{1}{\sqrt{2}}b(ud-du) \), \( \Sigma_c^{++} = cuu \), and \( \Sigma_c^+ = \frac{1}{\sqrt{2}}c(ud+du) \), we obtain the overlap,
\begin{align}\label{wmissioneq}
    \langle cdu~|~\frac{1}{\sqrt{2}}c(ud+du)~\rangle = \frac{1}{\sqrt{2}}.
\end{align}
Hence the decay mechanism, involving internal emission, is depicted in Fig.~\ref{feynDiag_Wission}. 

The same topology as in Fig.~\ref{feynDiag_Wission} (a) stands for the production of $\Sigma_c^{++} D^{*-} K^{*-}$.
\subsection{\( \Lambda_b \to \Sigma_c^{++} D^- K^- \),~~ $ D^- K^- $ mass distribution}

\begin{figure}[H]
  \centering
   \includegraphics[width=0.35\textwidth]{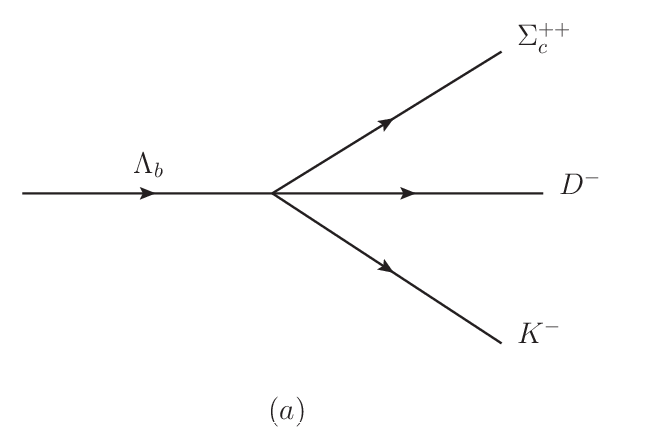}
   \includegraphics[width=0.6\textwidth]{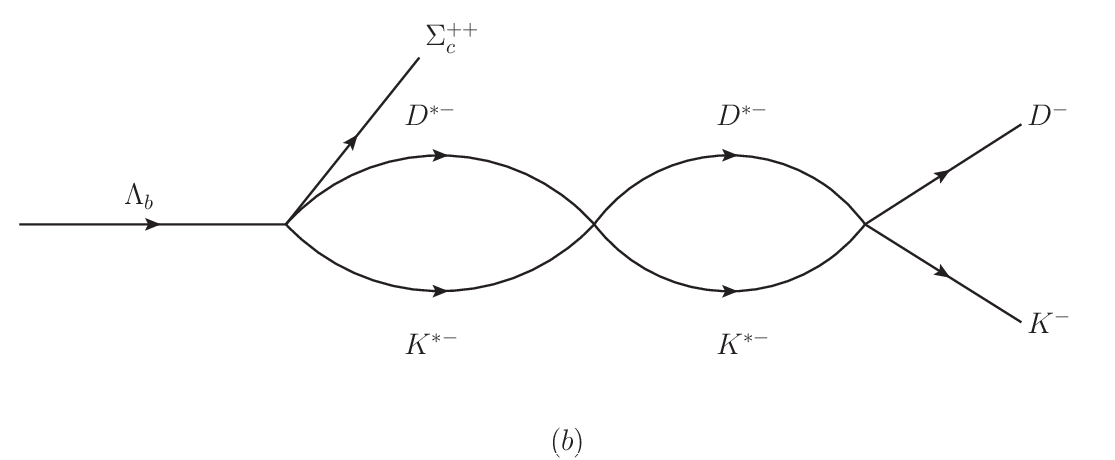}
  \caption{Mechanism of the reaction \( \Lambda_b \to \Sigma_c^{++} D^- K^- \).}
   \label{feynDiag1_2}
\end{figure}

From the diagram of Fig.~\ref{feynDiag_Wission} (a), we get a direct production of $\Sigma_c^{++} D^- K^- $. There is another mechanism that contributes, when we produce $\Sigma_c^{++} D^{*-} K^{*-}$, and the $D^{*-} K^{*-}$ makes a transition to $D^- K^-$.
After the initial decay, the \( D^{*-} K^{*-} \) state interacts and undergoes transition into \( D^- K^- \), with isospin $I=1$, and quantum numbers $J^P = 2^+$ of the resonance.  This process is illustrated in Fig.~\ref{feynDiag1_2}.
The mechanism of Fig.~\ref{feynDiag1_2} (a), tree level, can proceed via $S-$wave.
On the other hand, the $D^{*-}K^{*-}$ state produced in Fig.~\ref{feynDiag1_2} (b), with {orbital} angular momentum zero and total spin $J=2$, decays to $D^- K^-$, which implies that this pair has $L=2$ to conserve spin and parity. This produces an amplitude proportional to $Y_{20}(\theta)$, with $\theta$ the angle of $K^-$ with respect to $\Sigma_c^{++}$ in the rest frame of $D^- K^-$ (see details in the similar $B^+\to D^{+}D^-K^+$ reaction studied in Ref.~\cite{Bayar:2022wbx}). 
The transition amplitude \( t \) for Fig.~\ref{feynDiag1_2} can then be expressed as:
\begin{align}\label{T_DK}
& t = \tilde{a}\frac{\tilde{k}}{M_{\Lambda_b}}Y_{00} + b^{'}\frac{\tilde{k}^2}{M_{\text{inv}}^2(D^- K^-) - M_{R}^2 + i M_R \Gamma_R}Y_{20},
\end{align}
where $\tilde{k}$ is the momentum of the kaon in the $D^- K^-$ rest frame: 
\begin{align}\label{k_1}
& \tilde{k} = \frac{\lambda^{1 / 2}\left(M_{\text{inv}}^2(D^- K^-), m_{D^-}^2, m_{K^-}^2\right)}{2 M_{\text{inv}}(D^- K^-)}.
\end{align}
The values for $\tilde{a}$ and $b'$ from Ref.~\cite{Lyu:2024zdo} are 
 \( \tilde{a} = 13.48 \) and \( b^{'} = 1.38 \times 10^{-1} \). The mass of the resonance,  $M_R=2834$~MeV and the width  $\Gamma_R=19$~MeV, are taken from Ref.~\cite{Molina:2022jcd}.
Eq.~(\ref{T_DK}) contains an $S-$wave term in $D^- K^-$ corresponding to the tree level of Fig.~\ref{feynDiag1_2} (a), and a $D-$wave corresponding to  Fig.~\ref{feynDiag1_2} (b). In the absence of more concrete experimental information, we take it from the study of Ref.~\cite{Lyu:2024zdo} on the LHCb experimental reaction $B^-\to D^{*-}D^+K^+$~\cite{LHCb:2024vfz}. The $\tilde{k}$ factor in the $S-$wave $Y_{00}$ term is introduced {empirically} in Ref.~\cite{Lyu:2024zdo} to get the experimental background of Ref.~\cite{LHCb:2024vfz} and the $\tilde{k}^2$
 factor in the $D-$wave $Y_{20}$ term is introduced due to the $D-$wave character of the amplitude as in Ref.~\cite{Lyu:2024zdo}. Since in the experiment of Ref.~\cite{LHCb:2024vfz} there is no absolute normalization but just counts, the relevant magnitude is the relative weight of the $S-$wave to the $D-$wave, but the observation of this mass distribution in Ref.~\cite{LHCb:2024vfz}, with a similar weak decay pattern, proceeding also via internal emission, allows us to reasonably assume that this mass distribution is equally attainable.

The mass distribution for the decay \( \Lambda_b \to \Sigma_c^{++} D^- K^- \) is given by:
\begin{align}\label{1_Gamma}
    \frac{d\Gamma}{dM_{\text{inv}}(D^- K^-)~d\tilde{\Omega}} = \frac{1}{(2\pi)^4}\frac{1}{8M_{\Lambda_b}^2}p_{\Sigma_c^{++}}\tilde{k}\sum|t|^2,
\end{align}
where \( p_{\Sigma_c^{++}} \) is defined as:
\begin{align}
& p_{\Sigma_c^{++}} = \frac{\lambda^{1 / 2}\left(M_{\Lambda_b}^2, M_{\Sigma_c^{++}}^2, M_{\text{inv}}^2(D^- K^-)\right)}{2 M_{\Lambda_b}},
\end{align}
and \( \tilde{k} \) is given in Eq.~(\ref{k_1}), with \( \lambda \)  the standard Källén function.

Next, we define the moments of the angular distribution ($M_{\mathrm{inv}}\equiv M_{\text{inv}}(D^- K^-)$):
\begin{align}
& \frac{d \Gamma_l}{d M_{\mathrm{inv}}} = \int d \tilde{\Omega} \frac{d \Gamma}{d M_{\mathrm{inv}} d \tilde{\Omega}} Y_{l 0},
\end{align}
with $\tilde{\Omega}$ the solid angle in the $D^- K^-$ rest frame,
which, when applied to Eq.~\eqref{1_Gamma}, leads to the following relations,
\begin{align}\label{moments}
\frac{d \Gamma_0}{d M_{\mathrm{inv}}} & = FAC\left[|a|^2 + |b|^2\right], \\
\frac{d \Gamma_1}{d M_{\mathrm{inv}}} & = 0, \\
\frac{d \Gamma_2}{d M_{\mathrm{inv}}} & = FAC\left[\frac{2}{7} \sqrt{5}|b|^2 + 2 \text{Re}(ab^*)\right], \\
\frac{d \Gamma_3}{d M_{\mathrm{inv}}} & = 0, \\
\frac{d \Gamma_4}{d M_{\mathrm{inv}}} & = FAC \frac{6}{7} |b|^2.
\end{align}
Here, \( FAC \) is defined as:
\begin{align}
FAC = \frac{1}{\sqrt{4 \pi}} \frac{1}{(2\pi)^4}\frac{1}{8M_{\Lambda_b}^2}p_{\Sigma_c^{++}}\tilde{k}.
\end{align}
Thus, the total decay width is given by:
\begin{align}
\frac{d \Gamma}{d M_{\mathrm{inv}}} = \sqrt{4 \pi} \frac{d \Gamma_0}{d M_{\mathrm{inv}}}.
\end{align}
These results are consistent with the formalism presented in Refs.~\cite{Lyu:2024zdo,LHCb:2016lxy,Bayar:2022wbx}.

\subsection{\( \Lambda_b \to \Sigma_c^{++} D^- K^- \) $,~~\Sigma_c \bar{D} $ mass distribution}

According to the decay mechanism depicted in Fig.~\ref{feynDiag2_2}, plus the tree level as depicted in Fig.~\ref{feynDiag1_2} (a),
considering the diagrams of Fig.~\ref{feynDiag_Wission} and Eq.~(\ref{wmissioneq}), we can express the amplitude \( t \) as follows:
\begin{figure}[H]
  \centering
  \includegraphics[width=0.45\textwidth]{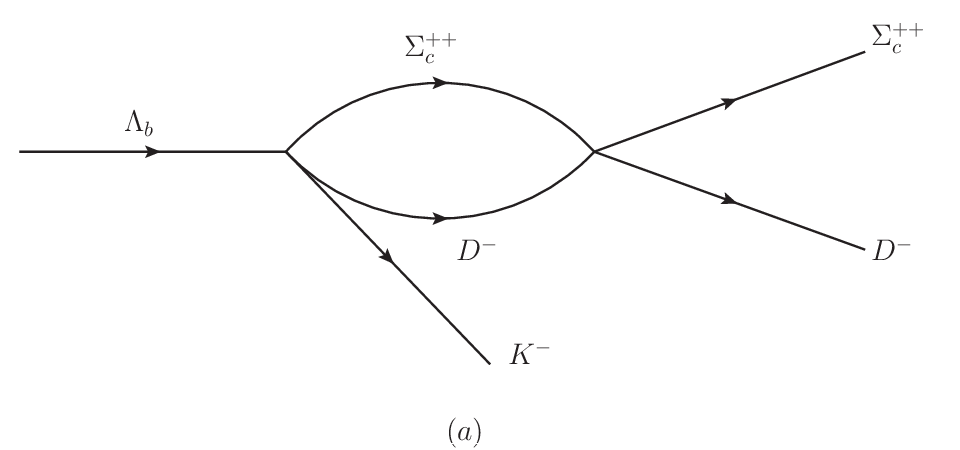}
  \includegraphics[width=0.48\textwidth]{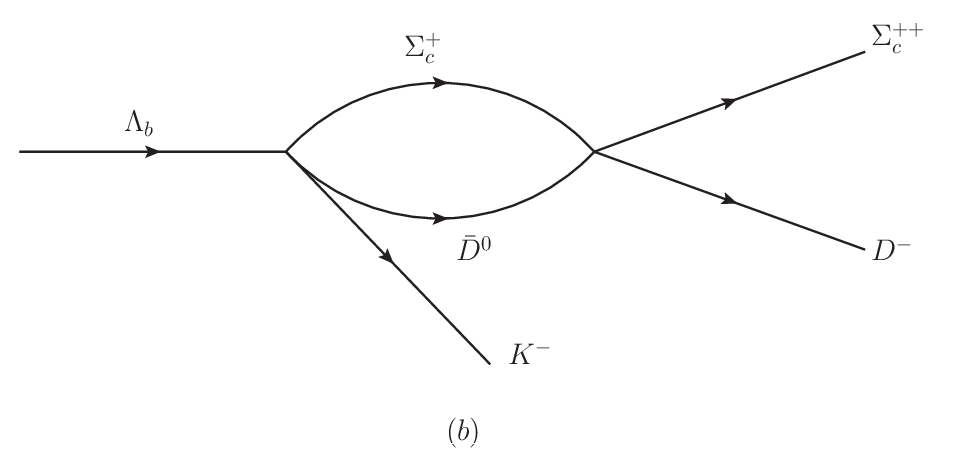}
  \caption{Feynman diagram for the reaction \( \Lambda_b \to \Sigma_c^{++} D^- K^- \). }
  \label{feynDiag2_2}
\end{figure}

\begin{align}\label{t_SbarD}
 t = \tilde{a} \frac{\tilde{k}'}{M_{\Lambda_b}} Y_{00} \Bigg[ 1 + G_{\Sigma_c\bar{D}}\Bigg(M_\text{inv}(\Sigma_c^{++} D^-)\Bigg)~t_{\Sigma_c^{++} D^-, \Sigma_c^{++} D^-}\Bigg(M_\text{inv}(\Sigma_c^{++} D^-)\Bigg)~~&\\\nonumber - \frac{1}{\sqrt{2}} G_{\Sigma_c\bar{D}}\Bigg(M_\text{inv}(\Sigma_c^{++} D^-)\Bigg)~t_{\Sigma_c^{+} \bar{D}^0, \Sigma_c^{++} D^-}\Bigg(M_\text{inv}(\Sigma_c^{++} D^-)\Bigg) \Bigg],
\end{align}
where the 1 corresponds to the tree level, and the second and third term to the diagrams of Fig.~\ref{feynDiag_Wission} (a) and Fig.~\ref{feynDiag_Wission} (b), respectively.
We include the global $\frac{1}{\sqrt{2}}$ factor of  $\Lambda_b = \frac{1}{\sqrt{2}}b(ud-du)$  in $\tilde{a}$.
The function $G$ is the loop function of  $\Sigma_c\bar{D}$ and $t_{ij}$ the transition amplitudes to be written below.
{ \( \tilde{k}' \) is the momentum of the \( K^- \) in the $D^-K^-$  rest frame,  \( \tilde{k} \),  which depends on the invariant mass of $D^- K^-$,} which is independent of the invariant mass of $\Sigma_c^{++} D^-$, but since we evaluate the $\Sigma_c^{++} D^-$ mass distribution close to threshold, we put the value of  \( \tilde{k} \)  at this threshold, which has a constant   value,  \( \tilde{k}' \). We make a derivation of this magnitude in Appendix A. We find
\begin{align}
& \tilde{k}' = \frac{\lambda^{1 / 2}\Bigg((m_{23})^2_{\max}, m_{D^-}^2, m_{K^-}^2\Bigg)}{2 (m_{23})_{\max}},
\end{align}
with $m_{23}$ the $D^- K^-$ invariant mass of $\Sigma_c^{++} D^-$ at  $M_\text{inv}(\Sigma_c^{++} D^-)=M_{\Sigma_c^{++} }+m_{D^-}$.
Concerning the \( \Sigma_c\bar{D} \) interaction in both the \( \Sigma_c^{++}D^- \) and \( \Sigma_c^{+}\bar{D}^0 \) channels, we can first write the isospin states,
\begin{align}
\vert \Sigma_c^{++} D^- \rangle &= \sqrt{\frac{1}{3}} \vert I = 3/2, I_3 = 1/2 \rangle + \sqrt{\frac{2}{3}} \vert I = 1/2, I_3 = 1/2 \rangle, \\
\vert \Sigma_c^{+} \bar{D}^0 \rangle &= \sqrt{\frac{2}{3}} \vert I = 3/2, I_3 = 1/2 \rangle - \sqrt{\frac{1}{3}} \vert I = 1/2, I_3 = 1/2 \rangle.
\end{align}
Hence, we obtain the  amplitude $t$, for $\Sigma_c\bar{D}$ interacting in $I=1/2$  as
\begin{align}
& t = \tilde{a} \frac{\tilde{k}'}{M_{\Lambda_b}} Y_{00} \left( 1 + G_{\Sigma_c\bar{D}}~t^{I=1/2}_{\Sigma_c\bar{D}, \Sigma_c\bar{D}}  \right).
\end{align}
The $\Sigma_c \bar{D}$ interaction is assumed to generate the $P_c(4312)$ resonance with isospin $I=1/2$, and the transition amplitude is written as:
\begin{align}\label{amplitudet2}
& t_{\Sigma_c\bar{D}, \Sigma_c\bar{D}}^{I=1/2} = \frac{g_{\Sigma_c\bar{D}}^2}{M_{\operatorname{inv}}(\Sigma_c\bar{D}) - M_{R} + \frac{i \Gamma_{R}}{2}},
\end{align}
where \( M_R \) and \( \Gamma_R \) represent the mass and width of the resonance, {$M_R=4312$~MeV, and $\Gamma_R=9.8$~MeV}~\cite{ParticleDataGroup:2024cfk}. The loop function $G_l$ in Eq.~(\ref{t_SbarD}) is given by:
\begin{align}\label{cut}
   G_l(\sqrt{s}) = \int_{|q|<q_{\text{max}}} \frac{d^3 q}{(2 \pi)^3} \frac{2 M_l \left( w_l(q) + E_l(q) \right)}{2 w_l(q) E_l(q)} \frac{1}{s - (w_l(q) + E_l(q))^2 + i \epsilon},
\end{align}
where \( M_l \) is the baryon mass, \( m_l \) is the meson mass, \( w_l(q) = \sqrt{m_l^2 + \mathbf{q}^2} \), and \( E_l(q) = \sqrt{M_l^2 + \mathbf{q}^2} \). The cutoff is \( q_{\text{max}} = 820~\text{MeV} \), as specified in Ref.~\cite{Xiao:2013yca}.

The mass distribution for the decay \( \Lambda_b \to \Sigma_c^{++} D^- K^- \) is represented by
\begin{align}\label{1_Gamma_Loop}
    \frac{d\Gamma}{dM_{\text{inv}}(\Sigma_c\bar{D})} = \frac{1}{(2\pi)^3} \frac{1}{4M_{\Lambda_b}^2} p_{K^-} \tilde{p}_{D^-} \sum |t|^2,
\end{align}
where \( \tilde{p}_{K^-} \) is the momentum of the \( K^- \) in the \( \Lambda_b \) rest frame and  \( \tilde{p}_{D^-} \)  the momentum of the \( D^- \) in the \( \Sigma_c\bar{D} \) rest frame. The momenta \( p_{K^-} \) and \( \tilde{p}_{D^-} \) are given by
\begin{align}
& p_{K^-} = \frac{\lambda^{1 / 2}\left(M_{\Lambda_b}^2, m_{K^-}^2, M_{\text{inv}}^2\left(\Sigma_c\bar{D}\right)\right)}{2 M_{\Lambda_b}}, \\
& \tilde{p}_{D^-} = \frac{\lambda^{1 / 2}\left(M_{\text{inv}}^2\left(\Sigma_c\bar{D}\right), M_{\Sigma_c}^2, m_{D^-}^2\right)}{2 M_{\text{inv}}\left(\Sigma_c\bar{D}\right)}.
\end{align}
The coupling $g_{\Sigma_c\bar{D}}$ in Eq.~(\ref{amplitudet2}) can be easily evaluated by using the {compositeness} condition with the normalization of fields in our approach, given in Ref.~\cite{danijuan} by 
\begin{align}
      g_{\Sigma_c\bar{D}}=\frac{E_{P_c}}{2M_{\Sigma_c}}\left(\frac{16\pi\gamma}{\mu}\right)^{1/2}; \qquad\qquad \gamma=\sqrt{2\mu B},  
\end{align}
where $E_{P_c}$ is the mass of $P_c(4312)$, $\mu$ the reduced mass of $\Sigma_c\bar{D}$ and $B$ the binding energy. We obtain
$$ g_{\Sigma_c\bar{D}}=2.25,$$   
that we use in this work. This coupling is not very different from the one obtained in Ref.~\cite{Xiao:2013yca}, $ g_{\Sigma_c\bar{D}}=3.12$, where a different binding was obtained theoretically.

\section{Results}

In this section, we present the calculated mass distributions for the decay $\Lambda_b \to \Sigma_c^{++} D^- K^-$, focusing on the systems $D^- K^-$ and $\Sigma_c \bar{D}$. The results are depicted in Figs.~\ref{res12_1} and \ref{res12_2}.

\subsection{Mass Distribution of \(D^-K^-\)}

The mass distribution for the \(D^- K^-\) system, as a function of the invariant mass \(M_{\text{inv}}(D^- K^-)\), is computed with the framework outlined in Section~II. The mass distribution shows a clear signal of the $2^+$ $\bar{D}^* \bar{K}^*$ state decaying into $D^- K^-$. We also computed various moments which provide information on the spin of the  \(D^- K^-\)  bound state.

\begin{figure}[H]
    \centering
\includegraphics[width=0.6\textwidth]{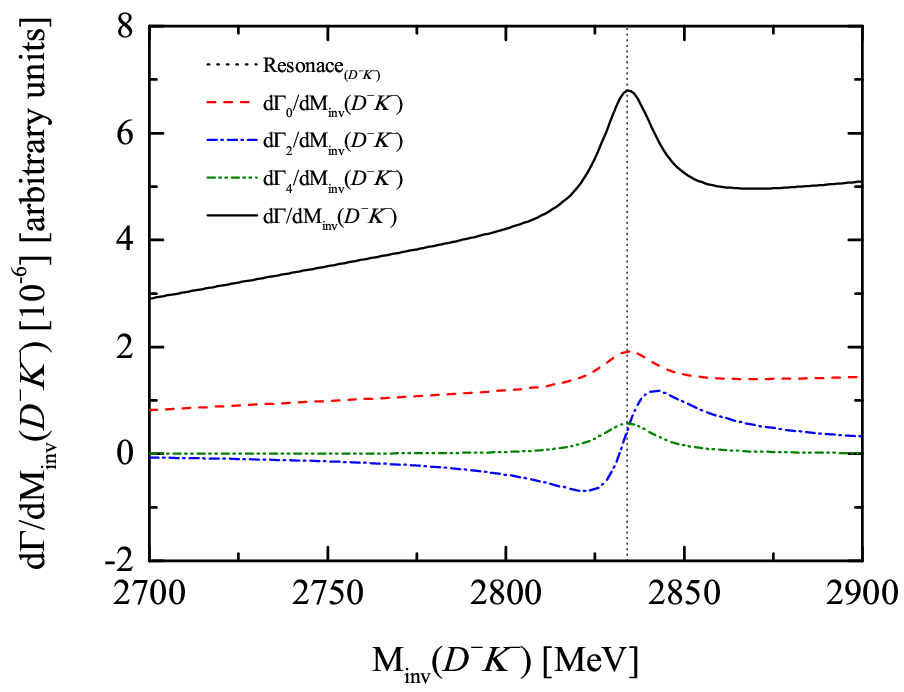}
   \caption{ The mass distribution $\frac{d\Gamma}{dM_{\mathrm{inv}}(D^-K^-)}$ and the different moments \(\frac{d\Gamma_i}{dM_{\mathrm{inv}}(D^- K^-)}\) of $D^-K^-$ as a function of $M_{\mathrm{inv}}(D^-K^-)$. The  vertical black dotted line  represents the position of $D^-K^-$ resonance.  }
    \label{res12_1}
\end{figure}

Figure~\ref{res12_1} illustrates the mass distribution \(\frac{d\Gamma}{dM_{\mathrm{inv}}(D^- K^-)}\) as a function of the invariant mass \(M_{\mathrm{inv}}(D^- K^-)\). The vertical black dotted line marks the resonance position of the \(D^- K^-\) system. Notably, the calculated mass distribution exhibits a significant peak at this resonance point. 
Using the input for $\tilde{a}$ and $b'$ from Ref.~\cite{Lyu:2024zdo}, we observe a clear signal of the $2^+$ state with respect to the background.
As we can see in Fig.~\ref{res12_1}, \(\frac{d\Gamma_4}{dM_{\mathrm{inv}}(D^- K^-)}\) shows a small peak resonance, actually $\frac{6}{7\sqrt{4\pi}}$  times the strength of the resonance over the background in \(\frac{d\Gamma}{dM_{\mathrm{inv}}(D^- K^-)}\), hence these two magnitudes provide the same information, only that in \(\frac{d\Gamma_4}{dM_{\mathrm{inv}}(D^- K^-)}\) the background has been eliminated. \(\frac{d\Gamma_0}{dM_{\mathrm{inv}}(D^- K^-)}\) also does not provide independent information since  it is $\frac{1}{\sqrt{4\pi}}$ the value of \(\frac{d\Gamma}{dM_{\mathrm{inv}}(D^- K^-)}\).
The relevant extra information is given by \(\frac{d\Gamma_2}{dM_{\mathrm{inv}}(D^- K^-)}\), which shows a distinct pattern, since we have now a change of sign of this magnitude at the resonance position rather than a peak. Also we benefit from the linear dependence of this magnitude in the resonance amplitude, and the strength measured from the minimum to the maximum is about four times bigger than the  strength at the peak of \(\frac{d\Gamma_4}{dM_{\mathrm{inv}}(D^- K^-)}\). We also find, in addition, that under our assumptions of the $S-$wave and $D-$wave alone, \(\frac{d\Gamma_1}{dM_{\mathrm{inv}}(D^- K^-)}\) and \(\frac{d\Gamma_3}{dM_{\mathrm{inv}}(D^- K^-)}\) are both zero. All this information is very valuable when it comes to determine the existence of a resonance and its spin.

\subsection{Mass Distribution of $\Sigma_c \bar{D}$}

\begin{figure}[H]
    \centering
\includegraphics[width=0.6\textwidth]{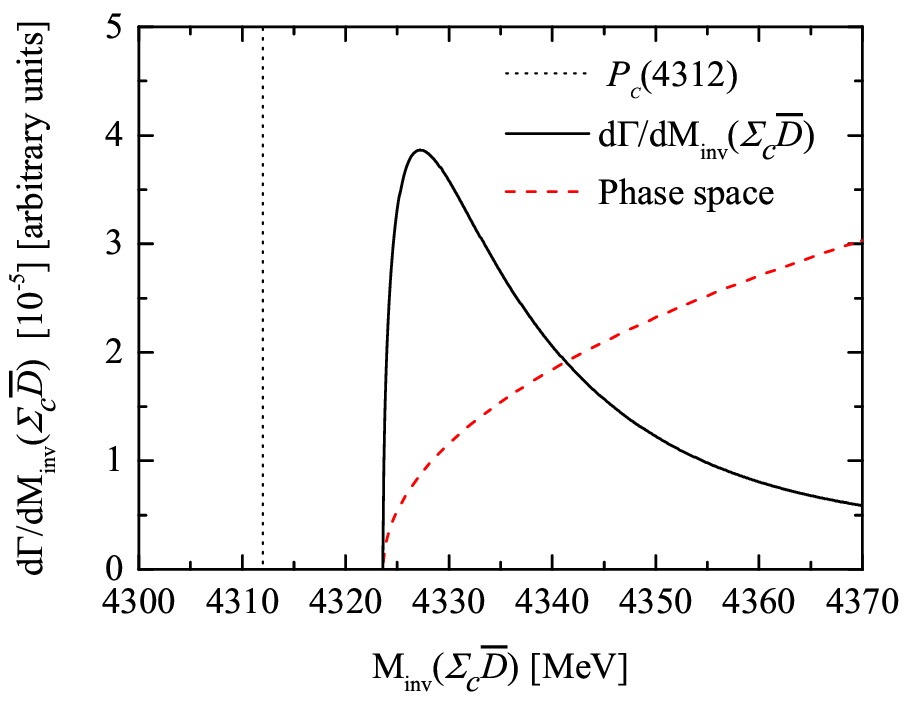}
   \caption{Mass distribution for $\Sigma_c \bar{D}$ as a function of $M_{\text{inv}}(\Sigma_c \bar{D})$ compared to phase space.  The vertical black dotted line  represents the $P_c(4312)$ mass. }
    \label{res12_2}
\end{figure}

Figure~\ref{res12_2} displays the mass distribution $\frac{d\Gamma}{dM_{\mathrm{inv}}(\Sigma_c \bar{D})}$ as a function of $M_{\mathrm{inv}}(\Sigma_c \bar{D})$.
In this figure, the vertical black dotted line indicates the  $P_c(4312)$ resonance. 
We observe a huge enhancement of the mass distribution close to threshold due to the presence of the $P_c(4312)$ state below threshold and its assumed molecular nature as a bound state of $\Sigma_c \bar{D}$. It is clear that if the $P_c(4312)$ state were not of molecular nature and had no link to the $\Sigma_c \bar{D}$ component we would not find any enhancement of the mass distribution at the $\Sigma_c \bar{D}$ threshold. The amount of the enhancement observed is thus linked to the $\Sigma_c \bar{D}$ component of the $P_c(4312)$ state. To make the enhancement obtained more apparent we have plotted the phase space distribution normalized to the same area. The difference in the shapes is unmistakable.

In summary, the results shown in Figs.~\ref{res12_1} and ~\ref{res12_2}, and the promise to learn about the existence of a new exotic state of $2^+$ and $D^*K^*$ nature, together with the information that we can obtain on the nature of the $P_c(4312)$ state, provide a clear incentive to measure these mass distributions as a continuation of present measurements of the branching ratio in the \(\Lambda_b \to \Sigma_c^{++} D^- K^-\) decay~\cite{LHCb:2024fel}.

\section{Conclusion}

In this study, we investigated the decay process \(\Lambda_b \to \Sigma_c^{++} D^- K^-\) with the goal of observing a molecular state of \(D^* K^*\)  molecular nature characterized by isospin \(I=1\),  along with quantum numbers \(J^P=2^+\), predicted theoretically.
 We computed the mass distribution of the \(D^- K^-\) system as a function of the invariant mass \(M_{\text{inv}}(D^- K^-)\) and analyzed various moments of this distribution. These moments revealed significant information about the $D^*K^*$ predicted bound state and its spin.

For the \(\Sigma_c \bar{D}\) mass distribution, due to the presence of the $P_c(4312)$ state below the \(\Sigma_c \bar{D}\)  threshold, assumed to be a  \(\Sigma_c \bar{D}\) bound state in our approach, we observed a prominent peak near the threshold region, which is substantially bigger than predictions based on phase space. 

Altogether, we found that the measurement of the \(D^- K^-\)  and   \( \Sigma_c^{++}D^- \)  mass distributions in the \( \Lambda_b \to \Sigma_c^{++} D^- K^- \) decay can provide much valuable information concerning the existence of a new exotic state of $J^P=2^+$ of \(D^* K^*\)  nature, while at the same time it can give us precious information about the nature of the $P_c(4312)$  state, a subject of permanent debate.

\section{Appendix A: Evaluation of $\tilde{k}'$ }
\begin{figure}[H]
  \centering
\includegraphics[width=0.4\textwidth]{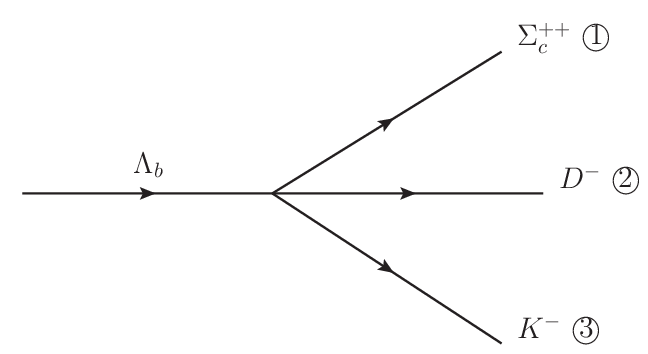}
  \caption{Feynman diagram for the reaction \( \Lambda_b \to \Sigma_c^{++} D^- K^- \). }
  \label{feynDiag2_2_lable}
\end{figure}

We label the particles as shown in Fig.~\ref{feynDiag2_2_lable}
From the Dalitz plot, using the formula from PDG~\cite{ParticleDataGroup:2024cfk}, the limits for \( m_{23}^2 \) are:
\begin{align}
& \left(m_{23}^2\right)_{\max} = \left(E_2^* + E_3^*\right)^2 - \left(\sqrt{E_2^{* 2} - m_2^2} - \sqrt{E_3^{* 2} - m_3^2}\right)^2, \\
& \left(m_{23}^2\right)_{\min} = \left(E_2^* + E_3^*\right)^2 - \left(\sqrt{E_2^{* 2} - m_2^2} + \sqrt{E_3^{* 2} - m_3^2}\right)^2,
\end{align}
where \( E_2^* \) and \( E_3^* \) are the energies of particles 2 and 3 in the 1, 2 rest frame, given by:
\begin{align}
& E_2^* = \frac{m_{12}^2 - m_1^2 + m_2^2}{2 m_{12}}, \quad E_3^* = \frac{M_{\Lambda_b}^2 - m_{12}^2 - m_3^2}{2 m_{12}}.
\end{align}

At the threshold of \( \Sigma_c^{++} D^- \), we have:
\begin{align}
& E_2^* = \frac{(M_{\Sigma_c^{++}} + m_{D^-})^2 - M_{\Sigma_c^{++}}^2 + m_{D^-}^2}{2 (M_{\Sigma_c^{++}} + m_{D^-})}= m_{D^-}, \\
& E_3^* = \frac{M_{\Lambda_b}^2 - (M_{\Sigma_c^{++}} + m_{D^-})^2 - m_{K^-}^2}{2 (M_{\Sigma_c^{++}} + m_{D^-})}.
\end{align}
Thus, the maximum and minimum values for \( m_{23}^2 \) are equal at the threshold:
\begin{align}
& (m_{23})^2_{\max} = (m_{23})^2_{\min} = m_{D^-}^2 + 2 m_{D^-} E_3^* + m_{K^-}^2,
\end{align}
and the momentum of the \( K^- \) in the \( D^-K^- \) rest frame is:
\begin{align}
& \tilde{k}' = \frac{\lambda^{1 / 2}\left((m_{23})^2_{\max}, m_{D^-}^2, m_{K^-}^2\right)}{2 (m_{23})_{\max}}.
\end{align}

\section{Acknowledgments}
This work is partly supported by the National Natural Science
Foundation of China under Grants  No. 12405089 and No. 12247108 and
the China Postdoctoral Science Foundation under Grant
No. 2022M720360 and No. 2022M720359. This work is also supported by
the Spanish Ministerio de Economia y Competitivi-
dad (MINECO) and European FEDER funds under
Contracts No. FIS2017-84038-C2-1-P B, PID2020-
112777GB-I00, and by Generalitat Valenciana under con-
tract PROMETEO/2020/023. This project has received
funding from the European Union Horizon 2020 research
and innovation programme under the program H2020-
INFRAIA-2018-1, grant agreement No. 824093 of the
STRONG-2020 project. This work is supported by the Spanish Ministerio de Ciencia e Innovaci\'on (MICINN) under contracts PID2020-112777GB-I00, PID2023-147458NB-C21 and CEX2023-001292-S; by Generalitat Valenciana under contracts PROMETEO/2020/023 and  CIPROM/2023/59.

\bibliography{refs.bib} 
\end{document}